# Influence of Absolute Humidity, Temperature and Population Density on COVID-19 Spread and Decay Durations: Multi-prefecture Study in Japan

Essam A. Rashed[1,2], Sachiko Kodera[1], Jose Gomez-Tames[1,3] and Akimasa Hirata[1,3,*]

[1]*Department of Electrical and Mechanical Engineering, Nagoya Institute of Technology, Nagoya 466-8555, Japan;*
[2]*Department of Mathematics, Faculty of Science, Suez Canal University, Ismailia 41522, Egypt*
[3]*Center of Biomedical Physics and Information Technology, Nagoya Institute of Technology, Nagoya 466-8555, Japan*
**\*** *Correspondence: ahirata@nitech.ac.jp Tel.: +81-52-735-7916 (A.H.)*

**Abstract:** This study analyzed the spread and decay durations of the COVID-19 pandemic in different prefectures of Japan. During the pandemic, affordable healthcare was widely available in Japan and the medical system did not suffer a collapse, making accurate comparisons between prefectures possible. For the 16 prefectures included in this study that had daily maximum confirmed cases exceeding ten, the number of daily confirmed cases follow bell-shape or log-normal distribution in most prefectures. A good correlation was observed between the spread and decay durations. However, some exceptions were observed in areas where travelers returned from foreign countries, which were defined as the origins of infection clusters. Excluding these prefectures, the population density was shown to be a major factor affecting the spread and decay patterns, with $R^2$=0.39 ($p$<0.05) and 0.42 ($p$<0.05), respectively, approximately corresponding to social distancing. The maximum absolute humidity was found to affect the decay duration normalized by the population density ($R^2$>0.36, $p$ <0.05). Our findings indicate that the estimated pandemic spread duration, based on the multivariate analysis of maximum absolute humidity, ambient temperature, and population density (adjusted $R^2$=0.53, $p$-value<0.05), could prove useful for intervention planning during potential future pandemics, including a second COVID-19 outbreak.

## 1. Introduction

Outbreaks of the infectious coronavirus disease of 2019 (COVID-19) have been reported worldwide [1,2]. In response, numerous studies on COVID-19 have been conducted on different aspects, including disease prevention, control, diagnosis, causes, and epidemiology [3]. A state of emergency was declared in 7 of Japan's 47 prefectures on April 7, 2020. This state of emergency was extended nationwide on April 16 and ended on May 25, 2020. During this state of emergency, people voluntarily self-isolated and no government-enforced lockdown of the prefectures (districts) was implemented.

Many epidemic modeling approaches exist [4] and various mathematical models have been used to demonstrate different aspects related to COVID-19, including serial intervals [5], transmission in China [6], and case studies in Wuhan, China [7], Italy [8], and Spain [9]. Comparisons among different countries have also been made [10] and the importance of different models in developing policies has been discussed in [11]. Regarding the COVID-19 spread in Italy, a multivariate analysis study was conducted to evaluate the effect of different environmental factors on confirmed positive cases [12]. Three regions in Italy with high national infection rates were considered. In a time-series analysis, different factors were evaluated within each region. Strong correlations were found with different environmental factors, such as the average temperature, humidity, and wind speed. Although this study addressed how these factors were correlated with confirmed COVID-19 cases, it does not discuss their influence on disease spread and decay durations. However, it does provide a parameter fitting model based on local data within each region. Being able to estimate the duration of the epidemic in different prefectures would be useful for developing lockdown policies.

The duration and morbidity rates of COVID-19, however, are not easy to compare because different co-factors affect them. The number of polymerase chain reaction (PCR) tests, a simple and cost-effective test, conducted in Japan was limited because of its reliability; therefore, chest computed tomography imaging was performed to obtain highly accurate diagnoses [13]. In addition, to avoid nosocomial infections and medical resource shortages, it was suggested that people with symptoms (e.g., fever > 37.5 °C for no more than four consecutive days) stay home and not seek immediate medical attention unless they had been in close contact with infected people or had recently visited a foreign country. Some patients have been reported to be asymptomatic [14], making the statistical study of COVID-19 more complex.

The statistics of confirmed deaths and COVID-19 positive patients are being updated every day in each prefecture of Japan. Furthermore, the Japanese medical system did not collapse during the outbreak, and the health insurance system provides COVID-19 medical care for free. However, the percentage of positive coronavirus tests ranges from 2.2 to 34.8% (Table 1) throughout Japan, making morbidity estimations in different areas more challenging. Assuming that the criteria for conducting a test are the same in each prefecture, the estimation of the epidemic duration provides a useful parameter.

Environmental co-factors potentially influencing COVID-19 morbidity, including ambient temperature, absolute humidity, and air pollution, have been studied [15-22]. However, neither these studies nor the above-mentioned modeling studies have considered the impact of population density and ambient conditions on both the spread and decay duration [23]. In fact, the durations were not discussed at all. The necessity of population density scaling in epidemic transmission models has been suggested for infectious diseases in general [24], indicating that the factors characterizing the spread and decay durations might be different. Recent studies have confirmed the effect of temperature and relative humidity on morbidity rates in Brazil [22,25] and the morbidity/mortality rates in Japan [26]. Studies with wider scopes include global data analysis—discussion on how temperature and humidity are correlated with the infection and fatality rates of the COVID-19 pandemic [27,28]. Both studies focus on the morbidity and mortality rates in their respective study area and globally, but neither of them discusses the effect of environmental factors on the duration of COVID-19, which is an essential factor for policymaking. Its effect on the mortality and morbidity rates is still controversial; for example, no correlation with weather data was found in a study of 122 cities in China [19] whereas a correlation was reported in [15]. A report highlighting the duration of the pandemic spread may lead to better management of medical resources and protective gear, which is a well-known life-threatening factor during the evolution of COVID-19.

In this study, we evaluate the effect of different ambient conditions and population densities on the spread and decay durations of COVID-19 in different prefectures of Japan. To the best of our knowledge, this is the first study to discuss the effect of environmental factors during COVID-19 in Japan. Japan provides an interesting case study for environmental factors, because medical services and social reactions are similar nationwide and high-quality data were properly recorded.

## 2. Materials and Methods

*2.1. Data Source*

Three datasets were utilized in this study. The first dataset contains the number of confirmed positive cases in each prefecture that was obtained from Toyo Keizai Online [29] and is based on a report by the Ministry of Health, Labour and Welfare [30].

Because the state of emergency was terminated on May 25, 2020, this study utilized data from March 15, 2020, to that date. COVID-19 started spreading in mid-February in the Aichi Prefecture, which is the only prefecture where the spread started earlier than March 15. As such, additional data were obtained from the Aichi Prefecture website [31] and included in this study.

The second dataset consists of population and prefecture area data that were obtained from the Statistics Bureau of Japan [32] (Table 1). Several prefectures are classified as geographically adjacent to primary prefectures (Tokyo, Osaka, Aichi, and Fukuoka; Table 1). In adjacent prefectures, several people were found to commute to the primary prefectures, thus potentially affecting the prefecture data, and should, therefore, not be neglected.

The third dataset contains weather data obtained from the Japan Meteorological Agency for each prefecture during the time of the pandemic. The absolute humidity was derived from the relative humidity and ambient temperature data [33].

**Table 1.** Population, area, and population density data of 19 prefectures where daily maximum cases of COVID-19 exceeded 10. Total cases of COVID-19 and mean percentages of the positive test cases are also listed.

|  | Population (×1,000) | Population density (people per km$^2$) | Total cases (through May 25) | Daily max cases | Percentage of positive test results (%) |
|---|---|---|---|---|---|
| Tokyo | 13,921 | 6,354.8 | 5,170 | 206 | 34.8 |
| Kanagawa* | 9,198 | 3,807.5 | 1,336 | 94 | 14.7 |
| Saitama* | 7,350 | 1,932.0 | 1,000 | 56 | 5.2 |
| Chiba* | 6,259 | 1,217.4 | 904 | 70 | 6.4 |
| Ibaragi | 2,860 | 470.4 | 168 | 28 | 3.7 |
| Gunma | 1,942 | 304.6 | 149 | 44 | 4.2 |
| Shizuoka | 3,644 | 467.9 | 75 | 18 | 2.2 |
| Aichi | 7,552 | 1,460.0 | 507 | 21 | 5.2 |
| Gifu** | 1,987 | 187.3 | 150 | 18 | 4.4 |
| Ishikawa | 1,138 | 271.7 | 296 | 20 | 11.2 |
| Toyama | 1,044 | 245.6 | 227 | 21 | 7.3 |
| Osaka | 8,809 | 4,631.0 | 1,781 | 108 | 6.1 |
| Hyogo*** | 5,466 | 650.4 | 699 | 57 | 6.4 |
| Kyoto*** | 2,583 | 560.1 | 358 | 20 | 4.6 |
| Shiga*** | 1,414 | 352.0 | 100 | 12 | 5.7 |
| Hiroshima | 2,804 | 331.1 | 167 | 51 | 2.5 |
| Fukuoka | 5,104 | 1,024.8 | 672 | 108 | 5.7 |
| Saga**** | 815 | 333.6 | 47 | 11 | 3.4 |
| Okinawa | 1,453 | 637.5 | 81 | 17 | 2.9 |

*Kanagawa, Saitama, and Chiba are considered adjacent prefectures of Tokyo.

**Gifu is considered an adjacent prefecture of Aichi.

***Hyogo, Kyoto, and Shiga are considered adjacent prefectures of Osaka.

****Saga is considered an adjacent prefecture of Fukuoka.

*2.2. Data Processing*

The number of confirmed positive tests may be influenced by diagnostic data, based on Japanese guidelines, serial intervals, and latency, which is affected by each doctor's judgment; therefore, the moving average over 7 days (±3 days in addition to the corresponding day) was considered to reduce the effect of potential singularities. This duration was used in the trajectory analysis of COVID-19 as well [34], also considering its incubation time (mean value of 5.1 days) [35]. Additional small peaks were still observed, which were attributed to clusters of patients (e.g., nosocomial infections, nursing home patients, and people returning from foreign countries). Similar to other countries, the number of confirmed positives may depend on the day of the week (see [34]); in general, the number of tests conducted on weekends is lower than that on weekdays.

This tendency prevents the direct fitting of the observed cases for comparison with a computational model (typically expressed as a bell-shape or log-normal distribution, similar to the trend in theoretical models [36,37]). Therefore, our study introduces the criteria of days required for spreading from 10% to 90% and for decaying from 90% to 10% of the peak [38] of the confirmed positives (7-day average [34]), as metrics for the evaluation, as shown in Fig. 1(a). This metric is particularly useful for cases where the noise effect should not be neglected. A limitation of this metric is that a certain minimum number of samples is needed because if the number of samples is small, one or two new patients would significantly influence the results. Therefore, as a pre-processing step, prefecture data were excluded if the daily maximum number of confirmed positive cases was less than 10. In addition, the percentage of morbidity/mortality in nosocomial infections and nursing home patients should not be significant. In Ishikawa, however, 75% of the confirmed deaths were nosocomial infections. Two prefectures (Saitama and Shiga) were excluded from the analysis due to a lack of humidity data, which are essential for discussing the effect of ambient conditions on the spread and decay durations (see the data source in the next subsection). Based on this criterion, 16 prefectures met the minimum number of confirmed positive cases, as shown in Fig. 1(b).

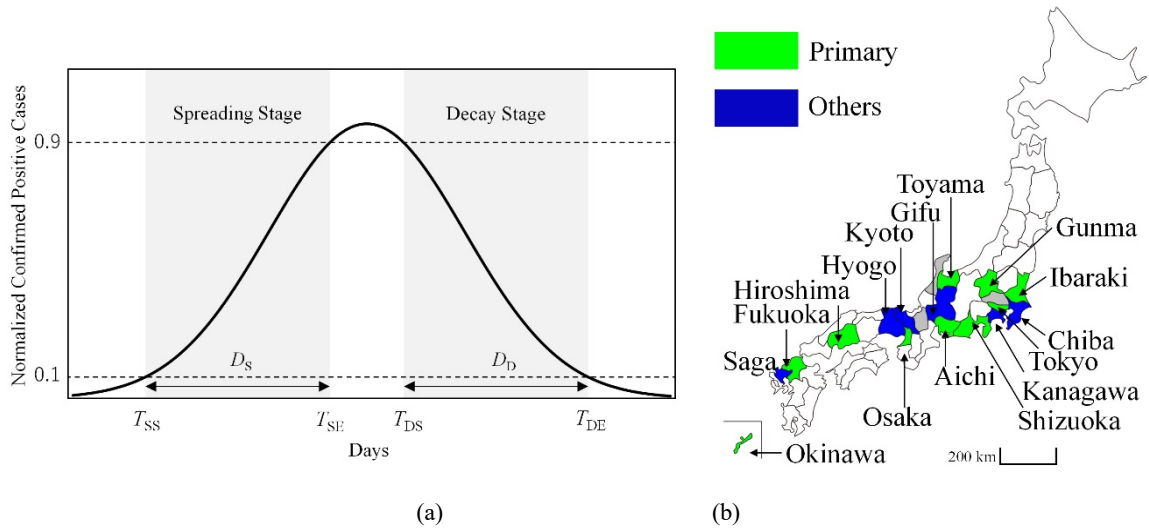

**Figure 1.** (a) Proposed definition of the spread and decay durations along the curve of the COVID-19 pandemic, which has been applied to the 7-day moving average of original data. Time parameters extracted from the definition are $T_{SS}$ (start of spread), $T_{SE}$ (end of spread), $T_{DS}$ (start of decay), $T_{DE}$ (end of decay), $D_S$ (spread duration), and $D_D$ (decay duration). (b) Map of Japan with 16 prefectures under study classified as primary prefectures and adjacent prefectures (others).

2.3. Statistics

A statistical analysis was conducted to determine the correlation between different factors and both the spread and decay periods of the pandemic using the software JMP (SAS Institute, Cary, NC, USA). To assist the pairwise correlations, we calculated the Spearman's rank correlation between the spread and decay durations normalized to the population density and different environmental parameters. The correlation matrix with partial correlation probabilities was also calculated. Subsequently, a multivariate analysis using linear regression was conducted using the same software. Statistical significance was accepted at $p<0.05$.

3. Results

Table 2 lists the starting and terminating dates of the spread and decay stages (using moving averages) as well as the date of the peak value of daily confirmed cases (without averaging) observed for each prefecture. Figure 2 shows the time-series of the confirmed new positive cases for the spread and decay durations based on the data presented in Table 2.

As shown in Fig. 2(a), the number of normalized confirmed cases increased in most prefectures, and multiple peaks are observed in some prefectures. In Aichi, the start of the spread period, $T_{SS}$, was three weeks earlier than in the other studied prefectures (Table 2), and in mid-March, when traveling residents returned home from foreign countries in response to a call from the Japanese government, two large peaks are visible in the spread graph. A similar tendency is observed in Saga, while in Chiba, and Gifu, only small peaks are observed.

As shown in Fig. 2(b), which illustrates the pandemic decay timeline, the number of normalized confirmed cases decreased over time in most prefectures. Similar to Fig. 2(a), multiple peaks are observed in some prefectures, including Kanagawa and Hyogo. In Aichi, the second peak, whose magnitude is approximately 0.9 times the maximum value, occurred on April15, 2020, and was caused by travelers returning from foreign countries [39].

Figure 3 illustrates the relationship between the durations of the spread and decay stages. These relationships are separated into two groups based on hierarchical clustering. The red group consists of metropolitan regions with a relatively high population density compared to the more rural blue group. Hereafter, the Aichi and Saga are considered to be outliers. Based on Fig. 3, we hypothesize that population density is an important factor, and further investigation is required.

Table 2. Starting and terminating dates of the spread and decay stages of the COVID-19 pandemic and the date (year 2020) when the highest daily peak value of confirmed cases was reported.

|  | $T_{SS}$ | $T_{SE}$ | Daily peak* | $T_{DS}$ | $T_{DE}$ | $D_S$ | $D_D$ |
|---|---|---|---|---|---|---|---|
| Tokyo | 17-Mar | 3-Apr | 17-Apr | 10-Apr | 7-May | 17 | 27 |
| Kanagawa | 19-Mar | 3-Apr | 10-Apr | 11-Apr | 19-May | 15 | 38 |
| Chiba | 19-Mar | 2-Apr | 17-Apr | 13-Apr | 5-May | 14 | 22 |
| Ibaraki | 16-Mar | 28-Mar | 3-Apr | 8-Apr | 23-Apr | 12 | 15 |
| Gunma | 25-Mar | 5-Apr | 11-Apr | 9-Apr | 22-Apr | 11 | 13 |
| Shizuoka | 25-Mar | 3-Apr | 10-Apr | 6-Apr | 27-Apr | 9 | 21 |
| Aichi | 22-Feb | 30-Mar | 4-Apr | 1-Apr | 27-Apr | 37 | 26 |
| Gifu | 25-Mar | 4-Apr | 8-Apr | 6-Apr | 17-Apr | 10 | 11 |
| Toyama | 1-Apr | 13-Apr | 17-Apr | 18-Apr | 30-Apr | 12 | 12 |
| Osaka | 18-Mar | 6-Apr | 14-Apr | 13-Apr | 6-May | 19 | 23 |
| Hyogo | 19-Mar | 4-Apr | 9-Apr | 7-Apr | 4-May | 16 | 27 |
| Kyoto | 16-Mar | 2-Apr | 7-Apr | 5-Apr | 9-May | 17 | 34 |
| Hiroshima | 26-Mar | 6-Apr | 12-Apr | 10-Apr | 27-Apr | 11 | 17 |
| Fukuoka | 22-Mar | 1-Apr | 11-Apr | 9-Apr | 27-Apr | 10 | 18 |
| Saga | 23-Mar | 15-Apr | 19-Apr | 22-Apr | 1-May | 23 | 9 |
| Okinawa | 28-Mar | 3-Apr | 7-Apr | 10-Apr | 25-Apr | 6 | 15 |

*Moving average is not applicable for daily peak values.

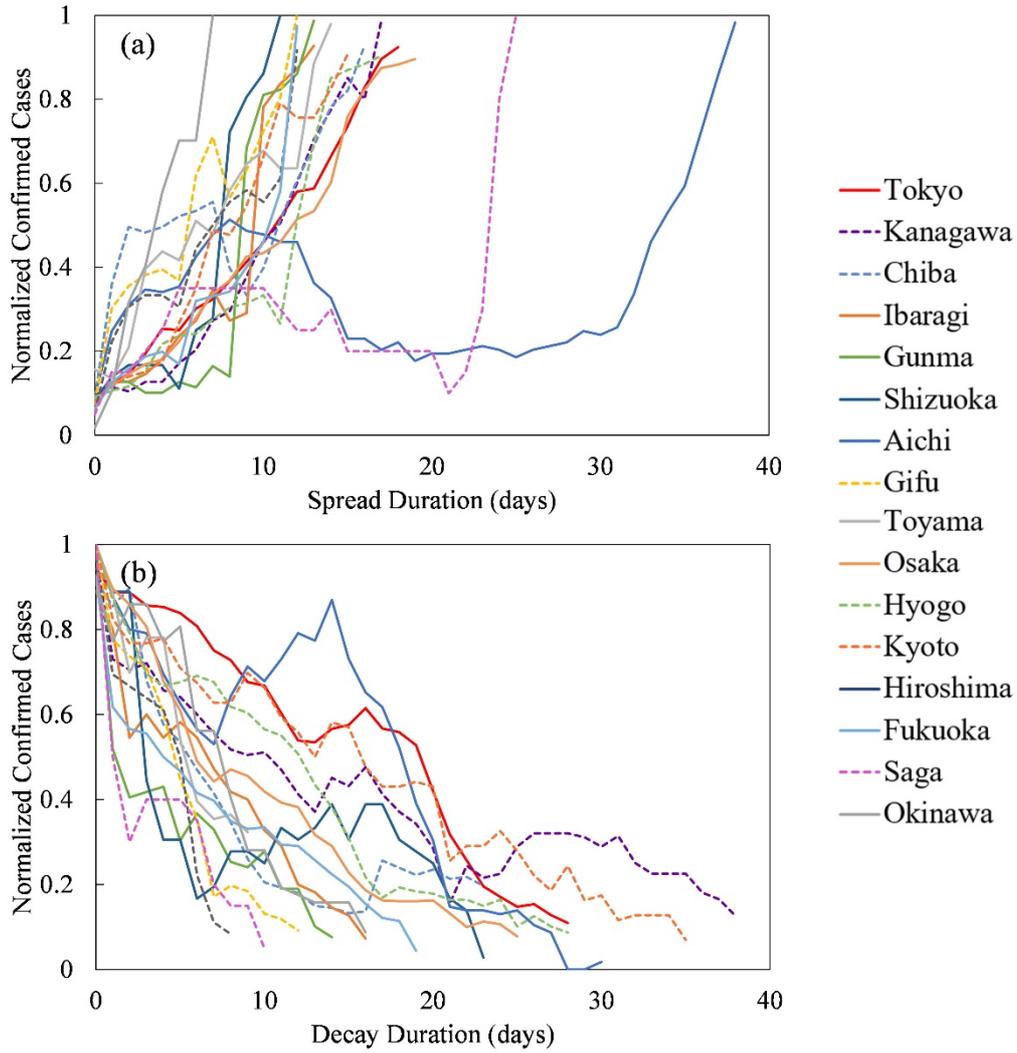

**Figure 2.** (a) Spread and (b) decay durations (in days) of 16 prefectures in Japan. Dashed lines represent the prefectures that are adjacent to primary prefectures (Tokyo, Osaka, Aichi, and Fukuoka).

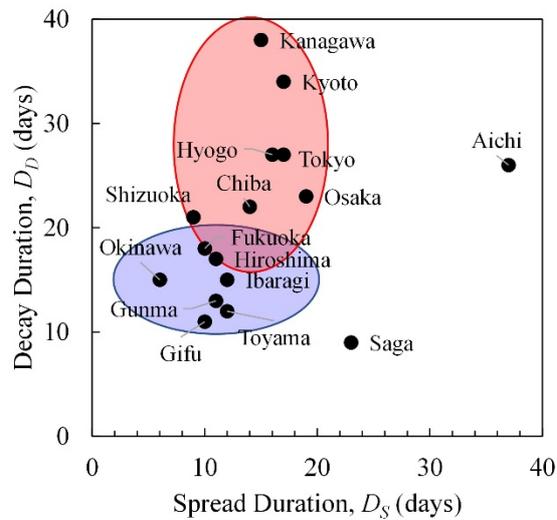

**Figure 3.** Relationship between the durations of spread and decay (in days). The red ellipse highlights prefectures with relatively high population densities, while the blue ellipse highlights those with lower population densities.

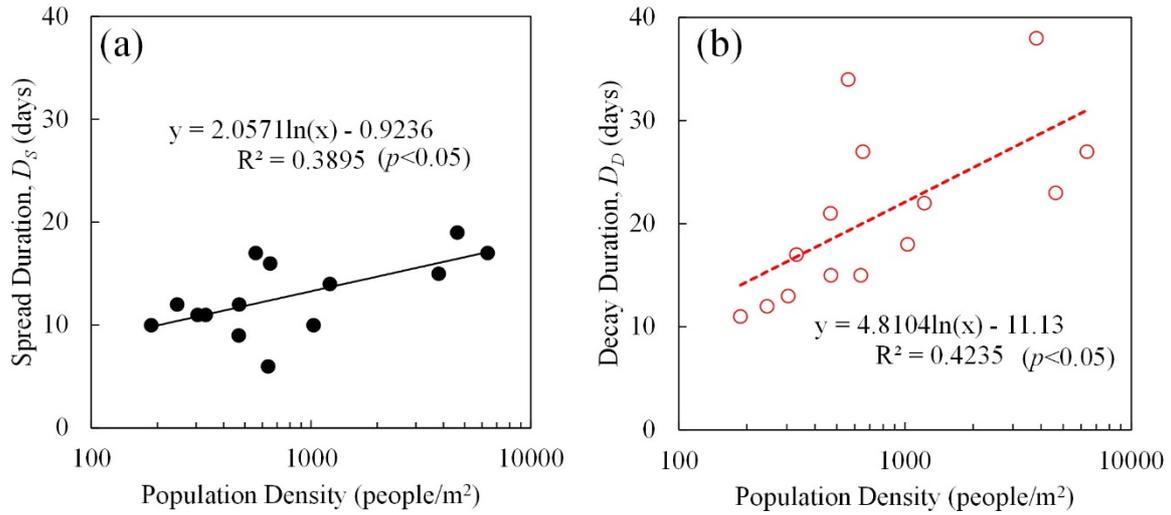

**Figure 4.** Relationship of the (a) spread and (b) decay durations ($D_S$ and $D_D$) with population density.

**Table 3.** Daily average, maximum and minimum temperature, and absolute humidity values. The values are averaged during the spread and decay stages as listed in Table 2. $T_{ave}$, $T_{max}$, and $T_{min}$ represent the daily average, maximum, and minimum temperatures, respectively. $H_{ave}$, $H_{max}$, and $H_{min}$ represent the daily average, maximum, and minimum absolute humidity values, respectively.

|   | $D_S$ | | | | | | $D_D$ | | | | | |
|---|---|---|---|---|---|---|---|---|---|---|---|---|
|   | $T_{ave}$ | $T_{max}$ | $T_{min}$ | $H_{ave}$ | $H_{max}$ | $H_{min}$ | $T_{ave}$ | $T_{max}$ | $T_{min}$ | $H_{ave}$ | $H_{max}$ | $H_{min}$ |
| Tokyo | 11.7 | 16.7 | 6.7 | 6.4 | 9.3 | 4.5 | 14.4 | 19.2 | 9.9 | 8.6 | 10.7 | 6.7 |
| Kanagawa | 12.4 | 16.7 | 8.0 | 6.8 | 9.7 | 4.7 | 16.6 | 20.7 | 13.0 | 9.8 | 11.7 | 7.7 |
| Chiba | 12.4 | 16.1 | 8.1 | 6.6 | 9.5 | 4.6 | 15.1 | 19.1 | 11.2 | 8.4 | 10.3 | 6.4 |
| Ibaragi | 10.3 | 17.1 | 3.4 | 5.7 | 8.5 | 3.7 | 10.8 | 15.6 | 6.4 | 6.5 | 8.2 | 4.9 |
| Gunma | 10.6 | 15.3 | 5.4 | 5.7 | 7.5 | 4.6 | 11.5 | 16.3 | 7.2 | 6.3 | 8.4 | 4.9 |
| Shizuoka | 13.1 | 16.6 | 9.3 | 8.6 | 10.6 | 6.6 | 14.3 | 18.7 | 10.0 | 7.1 | 8.9 | 5.4 |
| Aichi | 10.1 | 14.8 | 6.0 | 5.9 | 7.9 | 4.4 | 13.0 | 18.3 | 8.6 | 6.5 | 8.4 | 4.9 |
| Gifu | 12.0 | 16.4 | 7.7 | 6.7 | 8.4 | 4.9 | 12.6 | 18.2 | 7.7 | 5.1 | 6.6 | 3.6 |
| Toyama | 9.7 | 14.6 | 5.2 | 6.3 | 7.7 | 4.7 | 12.1 | 17.6 | 7.7 | 7.5 | 9.1 | 5.8 |
| Osaka | 12.7 | 17.0 | 8.9 | 6.7 | 8.9 | 5.1 | 16.2 | 20.6 | 12.3 | 8.1 | 10.2 | 6.3 |
| Hyogo | 12.7 | 16.4 | 9.1 | 7.2 | 9.6 | 5.3 | 15.5 | 19.0 | 12.4 | 8.1 | 9.5 | 6.0 |
| Kyoto | 11.5 | 16.6 | 6.8 | 6.4 | 8.6 | 4.7 | 14.7 | 20.1 | 10.0 | 7.1 | 9.0 | 5.3 |
| Hiroshima | 12.4 | 16.2 | 8.6 | 6.5 | 8.5 | 5.0 | 13.2 | 17.4 | 9.2 | 5.6 | 7.4 | 4.2 |
| Fukuoka | 14.2 | 17.5 | 11.3 | 8.8 | 11.0 | 6.9 | 14.0 | 17.5 | 10.9 | 7.3 | 9.4 | 5.7 |
| Saga | 13.4 | 17.9 | 9.0 | 7.3 | 9.1 | 5.5 | 14.9 | 20.1 | 9.8 | 7.1 | 8.6 | 5.4 |
| Okinawa | 21.3 | 24.0 | 18.8 | 14.7 | 17.4 | 12.4 | 19.8 | 22.1 | 17.6 | 11.8 | 14.1 | 10.0 |

Figure 4 shows the relationship between the spread and decay durations ($D_S$ and $D_D$) and the population density. The data of Aichi and Saga prefectures were excluded due to their outlier status. The resulting curves confirm the correlation between the spread and decay durations and the population density. The coefficients of determination of the spread and decay stages are 0.390 and 0.424 ($p<0.05$), respectively.

For environmental factors, the mean daily temperature and humidity values listed in Table 3 were obtained for the two durations listed in Table 2. Figure 5 shows the effect of the ambient parameters on the spread and decay durations. The daily mean, maximum, and diurnal change values were selected based on previous studies on the influenza virus [40,41] and suggest that the temperature and humidity variations are likely correlated with the COVID-19 mortality rate [15]. As seen in Fig. 5, such a correlation was observed for the data analyzed in this study. If the adjacent prefectures and those prefectures with multiple peaks in the curves are excluded, a mild correlation is observed for the maximum absolute humidity ($R^2 > 0.2$) with the spread and decay durations. In both cases, including and excluding adjacent prefectures, the maximum absolute humidity is the parameter that correlates significantly with duration. A brief discussion on how maximum ambient temperature and maximum absolute humidity are correlated can be found in Appendix A.

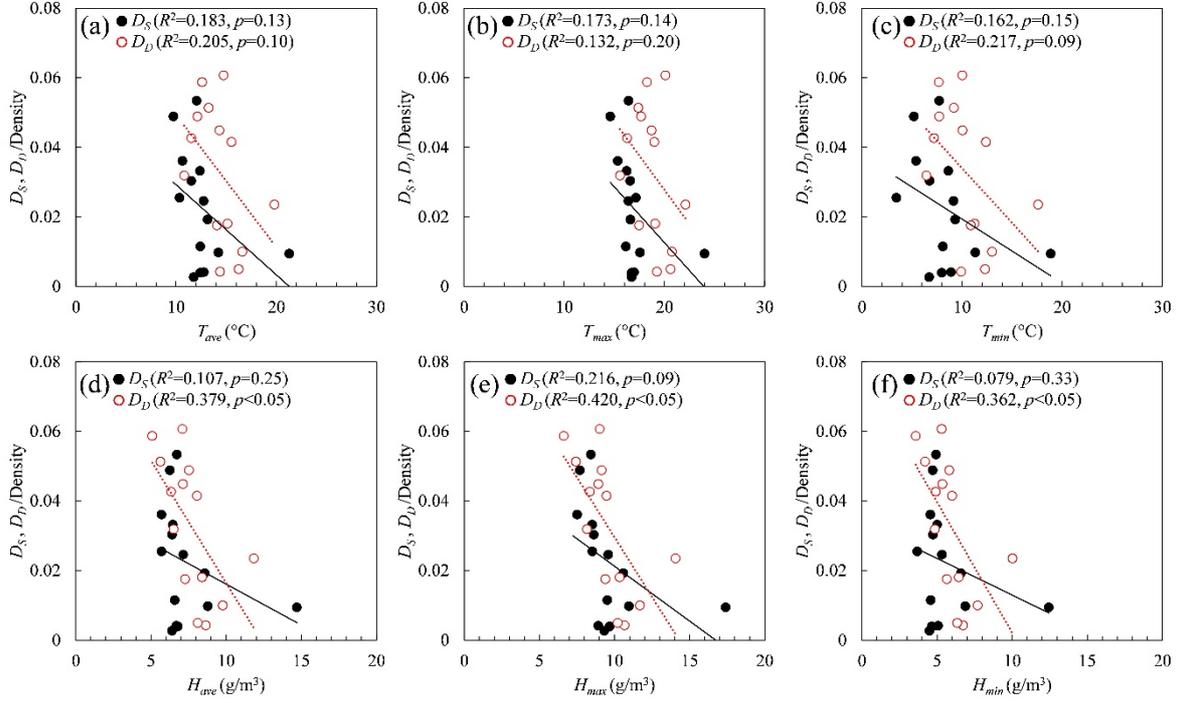

**Figure 5.** Relationship of the spread and decay durations ($D_S$ and $D_D$) with (a) daily average temperature ($T_{ave}$), (b) daily maximum temperature ($T_{max}$), (c) daily minimum temperature ($T_{min}$), (d) daily average absolute humidity ($H_{ave}$), (e) daily maximum absolute humidity ($H_{max}$), and (f) daily minimum absolute humidity ($H_{min}$).

**Table 4.** Spearman's rank correlation coefficients for spread and decay durations normalized by population density.

|  | Ds/Density | | Dd/Density | |
| --- | --- | --- | --- | --- |
|  | $\rho$ | $p$-value | $\rho$ | $p$-value |
| $T_{ave}$ | -0.526 | 0.05 | -0.459 | 0.099 |
| $T_{max}$ | -0.659 | **<0.05** | -0.385 | 0.175 |
| $T_{min}$ | -0.415 | 0.140 | -0.465 | 0.094 |
| $T_{diff}$ | 0.227 | 0.435 | 0.487 | 0.078 |
| $H_{ave}$ | -0.494 | 0.061 | -0.716 | **<0.05** |
| $H_{max}$ | -0.737 | **<0.05** | -0.741 | **<0.05** |
| $H_{min}$ | -0.130 | 0.657 | -0.733 | **<0.05** |
| $H_{diff}$ | -0.760 | **<0.05** | -0.718 | **<0.05** |

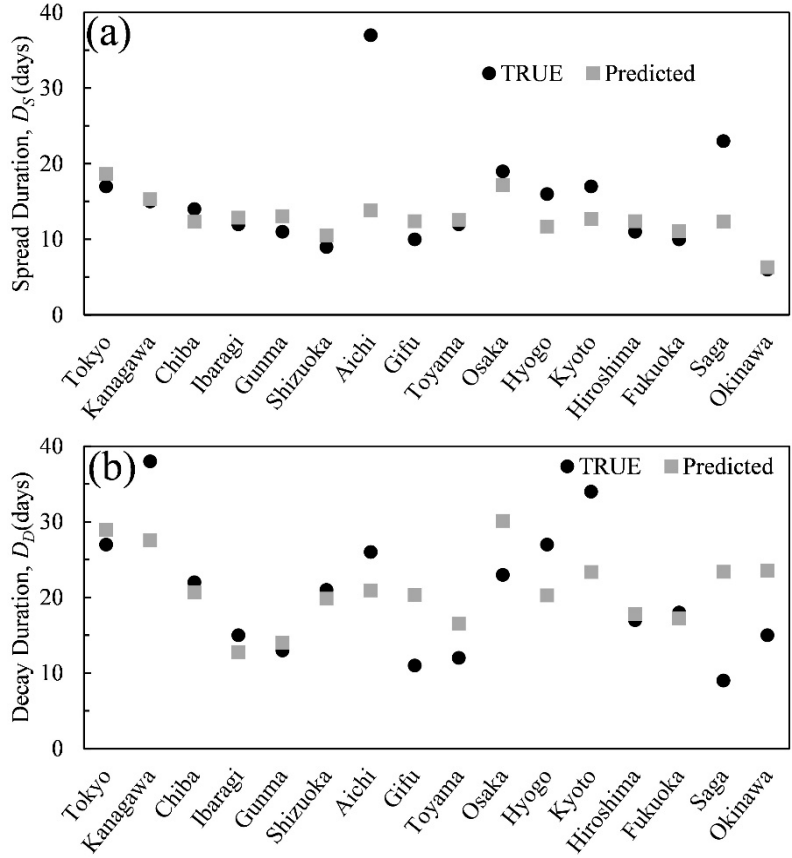

**Figure 6.** Multivariate regression (population density, maximum temperature, and absolute humidity) results for (a) spread and (b) decay periods.

**Table 5.** Coefficients of determination and adjusted $R^2$ values for the multivariate linear regression.

|   | $R^2$ | adj. $R^2$ | $p$-value |
|---|---|---|---|
| $D_S$ | 0.641 | 0.533 | **<0.05** |
| $D_D$ | 0.416 | 0.240 | 0.130 |

A multivariate correlation analysis was conducted to examine the relationship between the daily maximum temperature and absolute humidity values measured during the study period (pandemic spread and decay) and the population density (Fig. 6). The data of Aichi and Saga prefectures were excluded due to their outlier status. The results yield adjusted $R^2$ values of 0.53 ($p<0.05$) and 0.24 ($p=0.130$) in the spread and decay stages, respectively, indicating variability within some prefectures. The standardized partial regression coefficients of population density, daily maximum temperature, and absolute humidity are 0.611, 0.203, and -0.682, respectively, for the spread duration, while they are 0.388, 0.544, and -0.219, respectively, for the decay duration (Fig. 6).

## 4. Discussion

The effect of temperature and humidity on morbidity and mortality has long been investigated in research on health problems caused by different environmental factors (e.g., [42,43]). Even for COVID-19, as reviewed in the introduction, their influence has been discussed extensively in different countries. A case study of COVID-19 in Japan highlights the environmental factors in view of the consistency of data recording, healthcare quality, and social response to the outbreak of the disease. The limitation of data, which is partly attributable to different policies in Japan, as in Section 2, is one factor hampering the analysis of available COVID-19 data. A further challenge is the large variation in data reliability, consistency, and uniformity.

In this study, the durations of the spread and decay stages were evaluated, for the first time, to assist with the setting of policies and action plans. Although the results obtained in this study cannot be directly implemented in all cases, it provides valuable insight into how environmental factors influence the transmission phase during a pandemic in a more general framework.

The COVID-19 spread and decay durations of 16 prefectures in Japan that had a daily maximum of confirmed positive cases exceeding 10 were compared. A definition of the metrics for both durations from 10–90% of the normalized confirmed cases has been introduced. Normalization was performed to avoid potential differences in the absolute number of cases reported among prefectures due to different regulations. One difficulty in handling the data was the lower number of cases in Japan compared to most other countries; for example, the number is one to two orders of magnitude smaller than that of most European countries (e.g., see [34]). As mentioned above, a time-series analysis of COVID-19 cases in Japan is not feasible due to the Japanese COVID-19 policy combined with the incubation period.

The results of this study show that the number of confirmed cases generally increased and then decreased, forming bell-shaped or log-normal curves [36,37]. However, significant dual peaks are observed in Aichi and Saga during the spread stage and in Aichi, Saga and Tokyo during the decay stage. According to reports of each prefecture, this is mainly caused by patient clusters, which are attributed to nosocomial infections, nursing home patients, and travelers returning from foreign countries. In Aichi, at least one peak is caused predominantly by individuals who returned from foreign countries. The duration of such peaks is relatively short compared to the prefecture durations and generally shorter than the estimated ones. The total spread and decay durations in Aichi were shorter than 20 days for the first bell-shaped curve, whereas the estimated duration from the multivariate analysis was 35 days and thus, may not be considered as a second outbreak. If the duration defining threshold is changed (e.g., 0.05. to 0.95), the tendency does not change for distributions following a bell-shape and the duration increases by 30%. However, some differences are apparent in prefectures where small peaks (i.e., cluster infections) are observed. In general, such prefectures have been excluded or treated separately, thus, not influencing our conclusions.

Most noteworthy is the evaluation of the effect of population density on the spread and decay durations. The results for the primary prefectures show a good correlation between these parameters, indicating that population density functions as a rough proxy for social distancing [44]. Our data indicate that it significantly influences the formation of the bell-shape of the curve describing the pandemic. By normalizing our data using population density, we found good correlations between the temperature and absolute humidity, and the identified durations. This is surprising because the ambient conditions in the real world vary significantly with time (e.g., [45-47]). Regarding ambient condition parameters, the daily maximum absolute humidity had the greatest effect.

The parameters obtained with multivariate linear regression analysis were applied for the predictions: adjusted $R^2$ values are 0.53 ($p<0.05$) and 0.24 ($p=0.13$) for the spread and decay durations, respectively; that is, the decay duration is not statistically significant. The small peaks observed in the prefectures adjacent to the primary prefectures (Fig. 2b) can be the potential reason for this.

This demonstrates that in adjacent prefectures the influence of the primary prefectures may not be neglected because residents commuting between the prefectures influence the spread and/or decay durations. For example, Hyogo prefectural governer decided to match the end of state emergency with that in Osaka; they are adjacent with each other.

Therefore, additional constraints were defined to exclude prefectures located adjacent to primary prefectures to avoid the influence of infections caused by neighboring major pandemic clusters. A set of nine prefectures demonstrated a strong correlation of the average absolute humidity and population density, yielding adjusted $R^2$ values of 0.896 ($p<0.05$) and 0.681 ($p<0.05$) in the spread and decay stages, respectively.

The importance of absolute humidity was suggested in previous studies on morbidity/mortality rate of the influenza virus [48,49] and also in a preliminary study of COVID-19 [22]. Note that, the differences in other parameters, including maximum and average temperatures, and average absolute humidity, show comparable correlations. In [22] the role of the ambient parameters could not be adequately constrained, and controversies and inconsistencies remained (see the review in "Introduction").

A straightforward comparison of our findings with those of different countries is not feasible because the data is measured based on different closure policies, testing rates, validation measures, etc., which would introduce several biases. Further factors influencing data accuracy are lockdown restrictions, social distancing, and public response. Recently, the effect of air-conditioning units on COVID-19 transmission has also been suggested [50]. This case study on the pandemic duration in Japan nevertheless provides a highly consistent result that may lead to a better understanding of this correlation.

## 5. Conclusions

This study investigates the correlation of the durations of the spread and decay stages of COVID-19 with ambient conditions and population density in 16 prefectures of Japan. Although the number of cases reported in Japan is limited, the uniformity of the data provides a fairly consistent conclusion; the population density is dominant, and higher absolute humidity and temperature may result in shorter durations. Some exceptions exist, especially in prefectures where travelers returned from foreign countries. In addition, the durations in adjacent prefectures were influenced by the primary prefectures (e.g., Tokyo, Osaka, Aichi, and Fukuoka), resulting in longer durations. For a potential pandemic, especially a potential second wave of COVID-19, this factor should be considered as well as the multi-city comparison for developing different protection policies.

## Appendix A

The relationship between the different ambient condition metrics and morbidity is often discussed (e.g., [51]). For the COVID-19 duration analyzed in this study, the correlation between the maximum temperature and maximum absolute humidity in the spread duration presented in Fig. A1 are significant ($p<0.05$). The duration of COVID-19 was generally limited to two months; thus, the metrics considered in the discussion of Fig. 6 are correlated unlike the discussion of their annual impact on influenza (e.g., [49]).

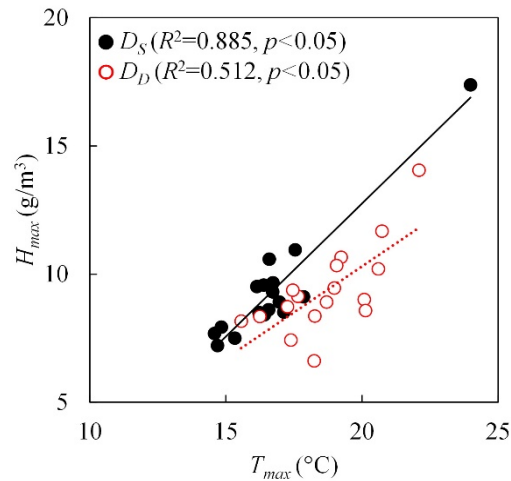

**Figure A1.** Relationship between maximum humidity and maximum temperature.